\documentclass[preprint,12pt,longtitle,authoryear]{elsarticle}
\usepackage{graphicx}
\usepackage[]{color}
\usepackage{epsfig}
\usepackage{latexsym}
\usepackage{amsmath}
\usepackage{graphicx}
\usepackage{amssymb}

\newtheorem{theorem}{Theorem}
\newtheorem{lemma}{Lemma}

\newtheorem{proposition}{Proposition}

\usepackage{geometry}
\usepackage{setspace}

\newcommand{\ignore}[1]{}

\newcommand{\PP}{\mathbb{P}}
\newcommand{\EE}{\mathbb{E}}

\newcommand{\blue}{\textcolor{black}}

\usepackage{amsmath}

\title{Dynamics of a \blue{birth--death  process based on combinatorial innovation}}
\author{Mike Steel$^a$, Wim Hordijk$^b$, and Stuart A. Kauffman$^c$}
\date{March 21, 2019}

\begin{document}

\begin{abstract}
A feature of human creativity is the ability to take a subset of existing items (e.g. objects,  ideas, or techniques) and combine them in various ways to give rise to new items, which, in turn,  fuel further growth. Occasionally, some of these  items may also  disappear (extinction).  We model this process by a simple stochastic birth--death model, with non-linear combinatorial terms in the growth coefficients to capture the propensity of subsets of items to give rise to new items. In its simplest form, this model involves just two parameters $(P, \alpha)$.  This process exhibits  a characteristic `hockey-stick' behaviour:  a long period of relatively little growth followed  by a relatively sudden `explosive' increase. We provide exact expressions for the mean and variance of this time to explosion and compare the results with simulations.  We then generalise our results to allow 
for more general parameter assignments, and consider possible applications to data involving human productivity and creativity. 
\end{abstract}

\bigskip

\address{ \textsuperscript{a} Biomathematics Research Centre, University of Canterbury, Christchurch, New Zealand.\\ {\em Email:} mike.steel@canterbury.ac.nz (corresponding author).}

\address{\textsuperscript{b} W. Hordijk:
Konrad Lorenz Institute for Evolution and Cognition Research, \\Klosterneuburg, Austria 
{\em Email:} wim@worldwidewanderings.net}

\address{\textsuperscript{c} S. Kauffman:
Institute for Systems Biology, Seattle, WA, USA 
{\em Email:} stukauffman@gmail.com}

\maketitle


\newpage 


 
\section{Introduction}

In this paper, we  introduce and analyse a new mathematical model for the broad process of the cumulative, combinatorial nature of technological evolution \citep{art, kau08, kau16, kau19, ogb, rea, val}.   Consider the course of  technological evolution over the course of hominid evolution \citep{str16, str17}. In the Lower Paleolithic, about 2.6 million years ago, our ancestor {\it Australopithecus} first started shaping simple stone tools such as diggers and scrapers. The diversity of these tools was perhaps a dozen or so. During the Upper Paleolithic and Mesolithic, the diversity of stone tools increased at a glacial pace. Stone knife blades grew longer and sharper over hundreds of thousands of years. Compound tools such as a knife blade hafted to a bone handle by sinew emerged perhaps 300,000 years ago. {\it Homo sapiens} arose about 150,000 years ago. By the time of Cro Magnon, 50,000 to 15,000 years ago, the number of stone tools had increased to perhaps several hundred, ranging from bone needles and bone flutes to arrow heads, fluted fish hooks, and the spear thrower. Ten thousand years later, in the time of Mesopotamia about 5000 years ago, the number of tools had increased to perhaps thousands, ranging in complexity from needles and pots to war chariots. 

Still another 5000 years later (today),  the diversity of `tools' has exploded into the billions, ranging from the 
60,000-year-old needle to machine tools, televisions, computers, and the International Space Station (ISS).  The Wright Brothers took their first flight in 1903, and a mere 66 years later, Apollo landed on the moon. 

This history of human technological evolution shows two major features. At first, it proceeded at a glacial pace for a very long time as the \blue{complexity} of goods and tools increased very slowly. But then the process exploded upward, creating an enormous array of tools, from simple to complex. In this late explosion the rate of change increased enormously. In less than a century we have gone from the advent of computers to word processing, the World Wide Web, smartphones with thousands of apps, and the Space Shuttle, to name just a few.

In this paper, we analyse a simple model that can explain this initial long and slow advance, followed by a sudden `hockey-stick' upward trend in which \blue{ an increasing number of distinct items  of increasing complexity appear. Here increasing `complexity' means that newly-arising items (goods, tools etc) combine features of several existing items, which in turn have resulting from combining features of earlier items, and so on. Thus, although $M_t$ measures the number of distinct items, the growth of $M_t$ is associated with increasing complexity of the items themselves in the model presented here.}

This model \blue{develops the theme that ``combinatorics is at the heart of innovation [and so]  provides a possible rationale for the accelerating growth of innovations '' \cite{sol}.}  It is 
 based, in part on the notion of the `adjacent possible', which was introduced some time ago to refer to the new things that could possibly arise next, given what is in existence now \citep{kau08, kau19}. For example, before  the development of rocketry, the Space Shuttle and ISS could not arise. However, in the early 20th Century, when Robert H. Goddard was trying to invent rocketry, the space shuttle and ISS were already in the adjacent possible. What exists now does not necessarily cause, but certainly enables what could arise next.

\blue{Our model exhibits the phenomenon of an `explosion' within finite time due to a non-linear (positive feedback) terms. The phenomenon is well known,
both in the setting of deterministic differential equation modelling (see for example \cite{gor}), and stochastic birth-processes (see e.g. \cite{feller, norris}).  Our emphasis here is to establish results particular to our model (e.g. expressions for the expected time to extinction) which do not directly follow from more general results.}

\subsection*{Formalising a model of \blue{combinatorial innovation}}

We propose that a simple cumulative combinatorial process underlies this pattern of human technological evolution. Humans take whatever lies at hand to fit a purpose and combine these in different possible ways, seeking combinations of  them that might together serve the desired purpose. These (possibly arbitrary) combinations are then tested to see if any of the new artifacts work, thus accumulating new goods or tools that are useful in some way.

This simple feature of human inventive exploration suggests the following equation (from \cite{kop18}):
\begin{equation}
\label{meq}
M_{t+1}  = M_t  + \sum_{i=1}^{M_t} \alpha_i \binom{M_t}{i},
\end{equation}
where $M_t$ is the number of goods or tools in the economy at time $t$ and  $\alpha_i, i \geq 1$ is a decreasing sequence of positive real numbers (each less than 1.0) which reflects the decreasing ease of finding and testing useful combinations among an increasing number of goods. 

However, Eqn.~(\ref{meq}) has a number of shortcomings.  Firstly, it will generally require $M_t$ to  take non-integer values, 
which is problematic for interpreting both the term $\binom{M_t}{i}$ and the range of summation\footnote{Rounding $M_t$ down to the nearest integer is one possibility.}.   Secondly, Eqn.~(\ref{meq}) is purely deterministic, whereas evolutionary processes are typically best modelled by a stochastic approach \blue{\citep{fel, yul}}. 
 Thirdly, Eqn.~(\ref{meq}) allows items to be gained but not lost.

In the next section, we describe and analyse a stochastic process, which we call the Combinatorial Formation (CF) model, based on Eqn.~(\ref{meq}), which avoids these shortcomings. In Section \ref{sec:simulation} we then show the results from two simulation models, one based on the deterministic Eqn.~(\ref{meq}) and one based on the stochastic CF model. These simulation results agree well with the theoretical predictions derived from the CF model. Moreover, the deterministic simulation model accurately represents the average behaviour of the stochastic simulation model. We end by discussing the implications of our model and its results for describing technological evolution, and how it could have (indirect) consequences for human evolution.

\section{A stochastic combinatorial formation (CF) model}  \label{sec:theory}

Consider the discrete-state, continuous-time process $M_t$ ($t\geq 0$) on the non-negative integers, describing the size of a population of `items'.  $M_0$ denotes the initial value of the process at time $t=0$. For each time $t>0$ consider the following Markovian transition process. Between time $t$ and $t+\delta$ (where $\delta$ is small):
\begin{itemize}
\item each non-empty subset $S$ of the population at time $t$ independently gives rise to a new item in the population with probability $\alpha_{|S|} \delta +o(\delta)$;
\item  each item in the population at time $t$ is independently removed from the population with probability $\mu \delta + o(\delta)$.
\end{itemize}
These two processes are assumed to proceed independently of each other in continuous time; in addition,  $o(\delta)$ refers to a term which is asymptotically negligible in proportion to $\delta$ as $\delta \rightarrow 0$. Observe that when $M_t=n$ the number of items added to the population in the interval between time $t$ and $t+\delta$ has a Poisson distribution with mean $(\sum_{i=1}^{n} \alpha_i \binom{n}{i})\delta +o(\delta)$,
whereas the number of items removed from the population in this interval has a Poisson distribution with mean $\mu n \delta =o(\delta)$.

The stochastic dynamics of $M_t$ can thus be described more concisely as follows. At time $t+\delta$:
$$M_{t+\delta} = M_t + \chi(M_t),$$
where, conditional on $M_t=n$ we have:
$$
\chi(M_t) = \begin{cases} 
+1, & \mbox{with probability } \delta  \sum_{i=1}^{n}\alpha_i \binom{n}{i} + o(\delta);\\
-1, & \mbox{with probability }   \delta \mu n+ o(\delta);\\
0, & \mbox{with probability 1 minus the sum of the other two probabilities}.
\end{cases}
$$

The value $\mu \geq 0$ is the rate at which individuals are removed from the system (i.e. an extinction event), which may depend on time $t$ (though we will mostly treat it as a constant, possibly zero). \blue{Extinction events are frequently observed in technological evolution, for example, the invention of the car marked the decline and fall of carriages, harness shops and buggies (see \cite{sol}).}

In the parlance of stochastic processes, $M_t$ describes a particular birth--death process, with a nonlinear (and time-independent) birth rate and a time-variable linear (or zero) death rate. What is slightly non-standard is that the range of the summation term in $\chi(M_t)$ (in the $+1$ case) depends on the random variable $M_t$.

The values $\alpha_1, \alpha_2, \ldots$ are non-negative constants.  We will assume throughout that they also satisfy the following condition:

\begin{equation}
\label{a1}
\alpha_{i}\neq 0 \mbox{ for some } i \leq  M_0, \mbox{ and } \alpha_k  \neq 0  \mbox{ for some } k\geq 2.
\end{equation}

The reason for imposing the first half of Condition (\ref{a1}) is that if $\alpha_i=0$ for all $i\leq M_0$, then $M_t$ either remains
constant for all time at $M_0$ (if $\mu=0$) or it is a pure death process (if $\mu>0$).  The reason for the second half of Condition~(\ref{a1}) is that if the largest value of $k$ for which $\alpha_k >0$ is $k=1$, then $M_t$ is described by a classic linear birth--death process (which behaves quite differently from the CF model).   We will refer to the value $k$ in Condition~(\ref{a1}) in some of the later proofs.

The following three quantities  play a key role in the dynamics of the CF model. Let: 
\begin{equation}
\label{lamu}
\lambda_n = \sum_{i=1}^n \alpha_i \binom{n}{i},\mbox{ } \lambda'_n = \lambda_n + \mu n, \mbox{ and } \gamma_n = \mu n/\lambda'_n.
\end{equation} 
The first quantity is the rate at which birth events occur, the second is the rate at which events (both birth or death)  occur, and the third is the probability  that when an event (birth or death) occurs, it is a death event. 

A particular instance of the CF model is the case where $\alpha_i = P \alpha^i$ for some $\alpha>0$   and $P \in (0, 1]$, which we refer to as the {\em geometric CF model}. Note that Condition (\ref{a1}) automatically holds in this case. Another special case is where $\mu=0$, in which case $M_t$ is described by a pure-birth process. We consider this special case first.

\subsection{The case when $\mu=0$}
\label{muzero}

In this case, with probability 1, there is a finite value $T$ (a random variable with a  finite mean and variance) for which $M_t$ tends to infinity as $t$ approaches $T$. Thus $T$ is the time until `explosion' of the process $M_t$. Our first  theorem provides an exact description of the mean and variance of the random variable $T$ for the pure-birth CF model.

\begin{theorem}
\label{mainthm}
\mbox{ } 
\begin{itemize}
\item[(i)] For the pure-birth CF model, the time to explosion ($T$) has a finite expected value and a finite variance given by: $$\EE[T] = \sum_{n=M_0}^\infty \lambda_n^{-1} \mbox{ and } Var[T] = \sum_{n=M_0}^\infty \lambda_n^{-2},$$
where $\lambda_n$ is as in (\ref{lamu}). 
\item[(ii)]
Consider now the geometric pure-birth CF model, and let $x=1+\alpha$, and $k= M_0 -1$. We have: 
\begin{equation}
\label{lamb}
\lambda_n = P\cdot ( x^n - 1),
\end{equation}
and so:
\begin{equation}
\label{lamb3}
\EE[T] = \frac{1}{P} \sum_{n >k } \frac{1}{x^n-1}.
\end{equation}
Moreover, a faster converging expression for $\EE[T]$ is given as follows:
\begin{equation}
\label{lamb2}
\EE[T] = \frac{1}{P} \cdot \left(\frac{x^{-k}}{x-1} + \frac{x^{-2k}}{x^2-1} + \frac{x^{-3k}}{x^3-1} + \cdots\right).
\end{equation}
Similarly,
\begin{equation}
\label{lamb4}
Var[T] = \frac{1}{P^2} \left(\frac{x^{-2k}}{x^2-1} + 2\cdot \frac{x^{-3k}}{x^3-1} + 3\cdot  \frac{x^{-4k}}{x^4-1} + \cdots \right).
\end{equation}
\end{itemize}
\end{theorem}

{\em Proof:}
  For Part (i), by the theory of continuous-time Markov processes \citep{gri}, $T$ is the sum of an infinite number of independent exponentially distributed random variables $(T_n: n \geq M_0)$, where $T_n$ has expected value given by $\EE[T_n] = \lambda_n^{-1}$ and thus variance $Var[T_n] = \lambda_n^{-2}$. Since the expected value (respectively, variance) of a sum of independent variables is the sum of the expected values (respectively variances), the equations stated
  for $\EE[T]$ and $Var[T]$ now follow.   It remains to show that these quantities are both finite.  To this end, observe that   
 Condition (\ref{a1}) implies  that
$\lambda_n \geq \alpha_i \binom{n}{i} + \alpha_k \binom{n}{k}$ where $\alpha_i, \alpha_k >0$ and $i \leq M_0$ and $k >1$. Thus, for $s=1,2$ we have:
$$\sum_{n=M_0}^\infty \lambda_n^{-s} \leq K\sum_{n=M_0}^\infty \left(\binom{n}{i}+\binom{n}{k}\right)^{-s},$$ for a constant  $K=1/(\min\{\alpha_i, \alpha_k\})^s$, and this infinite series has a convergent (finite) sum, as required. 

\bigskip

For  Part (ii), Eqn.~(\ref{lamb}) follows from the expression for $\EE[T]$ in Part (i), since  the Binomial Theorem gives $\sum_{i=1}^n P \cdot \alpha^i \binom{n}{i}= P\cdot ((1+\alpha)^n-1)$.

To establish Eqn.~(\ref{lamb2}), observe that:
$$\frac{1}{x^n-1} = \frac{1}{x^n(1-1/x^n)} = \frac{1}{x^n}\cdot \left(1-\frac{1}{x^n}\right)^{-1} = \frac{1}{x^n} \cdot \left(1+ \frac{1}{x^n} + \frac{1}{x^{2n}} + \cdots\right),$$
Thus, by Eqn.~(\ref{lamb3}) we have:
$$\EE[T] = \frac{1}{P} \sum_{n >k } \frac{1}{x^n} + \sum_{n >k } \frac{1}{x^{2n}} + \sum_{n >k } \frac{1}{x^{2n}} + \cdots.$$
Eqn.~(\ref{lamb2}) now follows, since: $$\sum_{n >k } \frac{1}{x^{jn}}  = \frac{(1/x^j)^{k+1}}{1-(1/x)^j} = \frac{x^{-jk}}{x^j -1},$$
for each value of $j \in \{1,2,3\ldots\}$. 

The expression for $Var[T]$ in (\ref{lamb4}) follows by a similar algebraic analysis to the expectation expression.
\hfill$\Box$

\bigskip

\noindent {\bf Example } 
\\
\\
Consider the pure-birth geometric CF model with $M_0=10$, $\alpha = 0.01$, and $P=1$. This gives $x=1.01$ and $k=9$. Summing the first 10 terms in the expression for $\EE[T]$ in Eqn.~(\ref{lamb2}) in Theorem~\ref{mainthm} gives a value of 219.47. The first 20 terms give 236.40, the first 50 terms give 241.58, and the first 100 (or more) terms gives $\sim$241.73. Similarly, the standard deviation of $T$ calculated by Eqn.~(\ref{lamb4}) is $\sigma=12.26$.

\ \\

\noindent {\bf Remarks}
\begin{itemize}
\item
The pure-birth CF model has a close connection to a classical process in population genetics. Let $H_n$ be the height of a Kingman coalescent tree, which traces the ancestry of $n$ genes back to their common ancestor \citep{wak}.  In the limit as $n \rightarrow \infty$, $H_n$ converges in distribution to the 
time to explosion $T$ for a pure-birth CF model that has $M_0=2, a_2\neq  0$ and $a_i = 0$ for all $i \neq 2$ (note that this is an instance of the CF model, as it satisfies Condition (\ref{a1})). In particular, $\EE[T] = \EE[H]=2$ and $Var[T] =Var[H]= (4\pi^2/3) - 12$ (these expressions for $E[H]$ and $Var[H]$ are classical coalescent results from the 1990s (\cite{wak}, p.76)).

\item
Theorem~\ref{mainthm} can be strengthened a little. For each integer $n \geq 1$, let $T^{(n)}$ be the time to explosion of a geometric pure-birth CF model conditional on $M_0=n$. Not only is $\EE[T^{(n)}]$ finite for each value of $n$ but $\sum_{n\geq 1} \EE[T^{(n)}]$ is also finite. This follows from the following identity:
\begin{equation}
\label{wow}
\sum_{n\geq 1} \EE[T^{(n)}] = \frac{1}{P} \sum_{j=1}^\infty \frac{x^j}{(x^j-1)^2},
\end{equation}
where (as before) $x=1+\alpha$. Eqn.~(\ref{wow}) follows from writing 
\begin{equation}
\label{wow2}
\sum_{n\geq 1} \EE[T^{(n)}]  = \frac{1}{P} \sum_{k=0}^\infty \left(\sum_{j=1}^\infty \frac{x^{-jk}}{x^j-1}\right)
\end{equation}
 (from Eqn.~(\ref{lamb2}) in  Theorem~\ref{mainthm}, noting that $k=n-1$) and then interchanging the order of summation
 in Eqn.~(\ref{wow2}).  The expression on the right of Eqn.~(\ref{wow}) is finite, since $\frac{x^j}{(x^j-1)^2} \leq \frac{1}{x^j -2}$, and the partial sums of these latter terms converge because $x>1$.

\end{itemize}

\subsection{The general CF model allowing extinction}

Let $X_n, n=1,2,3\ldots$ denote the (discrete time sampled)  value of $M_t$ at $t=n$.  We first remark that  process satisfies the following two conditions: 
\begin{itemize}
\item[(i)]  if $X_n=0$, then  $X_{n+1}=0$, and 
\item[(ii)] for some values $\delta_x>0$:
$$
X_n \leq x \Rightarrow \PP(\exists r: X_r=0|X_1, X_2, \ldots, X_n)\geq \delta_x.
$$
\end{itemize}
To see that (ii) holds, observe that since $X_n$ is a Markov process, we have:
$$\PP(\exists r: X_r=0|X_1, X_2, \ldots, X_n) = \PP(\exists r: X_r=0|X_n) \geq \PP(X_{n+1}=0|X_n).$$ 
Now, if $X_n\leq x$, then the probability that all the (at most $x$) items in $X_n$ are removed and no other items are added in the unit time interval from $n$ to $n+1$ is a strictly positive value that depends only on $x$ and the ($\mu, \alpha_i$)  parameters in the CF model.

A classic theorem of \cite{jag} states that for any process $X_n$ that satisfies conditions (i) and (ii) above,  the following holds: With probability 1, there is either a finite value of $n$ for which  $X_n$ equals zero (and so remains at zero),  or $X_n$ tends to infinity as $n \rightarrow \infty$. It follows that if  $M_t$ does not become extinct, it tends to infinity. However, tending to infinity is a different (weaker) condition than explosion (e.g. linear birth processes tend to infinity but do not explode)  and so we need to argue further for this. We do this by deriving a stronger result concerning the expected time until extinction or explosion, in Theorem 2.

\begin{theorem}
\label{thm2}
Consider the CF model.
\begin{itemize}
\item[(i)]
With probability 1, $M_t$ either explodes or becomes extinct.
\item[(ii)] The expected time until $M_t$ either explodes or becomes extinct is finite. 
\end{itemize}
\end{theorem}
{\em Proof:}
First observe that for the pure-birth CF model, extinction cannot occur and therefore Parts (i) and (ii) hold from the results in the previous section. Thus throughout this proof we will assume that $\mu>0$.

\bigskip

{\em Proof of Part (i):}   Consider the discrete-time sampled process $M_0, M_1, M_2, \ldots, M_i, \ldots, $
with $M_0 \neq 0$.    
We first establish the following claim:

{\bf Claim 1:} For a sufficiently large integer $m$, and all values of $n \geq m$, the probability that the discrete-sampled CF process has exploded prior to time $i+1$ (an event we denote by writing $M_{i+1}=\infty$)  conditional on $M_i=n$, together with the values of
$M_0, \ldots, M_{i-1}$, is $\geq p$
where $p>0$ is a value that depends only on $n$ and the $\alpha$ parameters and $\mu$.  In other words, for all $i\geq 0$:
\begin{equation}
\label{oneside}
\infty > n \geq m \Rightarrow \PP(M_{i+1} =\infty |M_i = n, M_0, \ldots, M_{i-1}, M_i = n) \geq p.
\end{equation}

We give a short proof of Claim 1 under the assumption that $\alpha_k>0$ for some $k \geq 3$ (which always holds in the geometric CF model). Claim 1 also holds when $\alpha_2>0$  and $\alpha_k=0$ for all $k>2$, but its proof requires a more delicate argument, given in the Appendix (essentially, if $M_t$ explodes, then the number of death events is finite in the case we deal with here, but in special the case dealt with in the Appendix it tends to infinity).

Select a sufficiently large value of $m$ so that the following two inequalities hold: 
\begin{equation}
\label{abab}
 \alpha_k\binom{m}{k} > \mu m  \mbox { and }  \sum_{n=m}^\infty \frac{1}{\lambda_n} <1,
 \end{equation} 
 where $\lambda_n$ is as given in Eqn.~(\ref{lamu}).
 The first inequality can clearly be satisfied (indeed it requires only that  $\alpha_k>0$ for some $k\geq 2$),  and the second inequality can be satisfied since  $\lambda_n$ is of order at most $n^{-2}$ (since $\alpha_k>0$ for some $k \geq 3$).  It follows that $\alpha_k\binom{n}{k} > \mu n$  for all $n \geq m$.  

Suppose that $M_{t'} = n$ where $n \geq m$. The probability that the first change in the value of $M_{t}$ after time $t'$ is a birth (rather than death)   event is exactly 
$\left( 1-\gamma_n \right)$ and this is at least  $\left(1-\frac{C}{(n-1) \cdots (n-k+1)}\right),$
where $C=k!\mu/\alpha_k$ (this follows from Eqn.~(\ref{lamu}), noting that $\gamma_n \leq \mu n/ \alpha_k\binom{n}{k}$. 
Thus, the probability that there are no death events after time $t'$ (conditional on  $M_{t'}=n \geq m$) is at least
$p= \prod_{i=m}^\infty\left(1-\frac{C}{(i-1) \cdots (i-k+1)}\right),$
which is strictly positive since $k\geq 3$.

However, conditional on the event that no deaths occur after $M_t=n$, the process $M_t$ is identical to a pure-birth CF process with the same $\alpha_i$ values (and $\mu=0$) with $M_t=m$.  Such a process has a strictly positive probability of exploding by time $i+1$ (here we use the fact that
$\PP(T\leq 1) \geq 1-\EE[T]$ and the right-hand side is strictly positive by Theorem 1(i) and the second half of the condition in ~(\ref{abab}).
 This establishes Claim 1.

\bigskip

Next observe that there is a value $p'>0$ (dependent only on $m$ and the CF parameters ($\alpha$ values and $\mu$) for which the following holds for all $i\geq 0$:
\begin{equation}
\label{twoside}
0<n < m \Rightarrow \PP(M_{i+1} =0|M_0, \ldots, M_{i-1}, M_i = n) \geq p'.
\end{equation}
To see this, simply observe that the process between $t$ and $t+1$ could begin with $i$ sequential deaths (and no births) and thus absorb at zero, all with strictly positive probability.

By combining (\ref{oneside}) and (\ref{twoside}), letting $p''= \min \{p, p'\}>0$, and letting $E_{i+1}$ be the event that $M_{i+1} = 0 \mbox{ or } M_{i+1}=\infty$ the following inequality holds for all values of $n$ (both $< m$ and $\geq m$)

\begin{equation}
\label{threeside}
\PP(E_{i+1} | M_0, \ldots, M_{i-1}, M_i=n) \geq p''.
\end{equation}

Let $\overline{E}_{i}$ denote the complement of $E_i$.  From (\ref{threeside}), we have:

\begin{equation}
\label{fourside}
\PP(\overline{E}_{i+1}|\overline{E}_{i}) \leq 1-p''.
\end{equation}
Since $\overline{E}_{i}$ is a nested decreasing sequence and $\overline{E_i}= \bigcap_{j=1}^i \overline{E}_{j}$, the product rule gives:
$$\PP(\overline{E}_{i}) = \PP(\overline{E}_{1})\cdot  \PP(\overline{E}_{2}|\overline{E}_{1})\cdot \PP(\overline{E_{3}}|\overline{E}_{2})  \cdots \PP(\overline{E}_{i}|\overline{E}_{i-1}).$$
Thus, from Inequality~(\ref{fourside}) we have:
$$\PP( \overline{E}_{i}) \leq (1-p'')^{i-1}.$$
It follows that $\sum_{i=1}^\infty \PP(\overline{E}_{i}) \leq 1/p'' < \infty$; therefore,  by the Borel--Cantelli Lemma, with probability 1 only finitely many of the events
$\overline{E}_{i}$ occur  \citep{gri}.  Thus, with probability 1, $M_i$ (and hence $M_t$) equals zero or explodes at a finite time.

\bigskip

{\em Proof of Part (ii):} 
If $T$ is the (continuous) time until $M_t$ first reaches 0 or explodes, then
$$T \leq \sum_{i=1}^ \infty 1_{\overline{E}_{i}},$$
where $1_{\overline{E}_{i}}$ is the indicator random variable that takes the value $1$ if $\overline{E}_{i}$ occurs, and 0 if $E_i$ occurs. 
Thus
$$\EE[T] \leq \EE\left[ \sum_{i=1}^\infty 1_{\overline{E}_{i}}\right] = \sum_{i=1}^\infty \EE[1_{\overline{E}_{i}}]  
=\sum_{i=1}^\infty
 \PP(\overline{E}_{i}) \leq \sum_{i=1}^\infty (1-p'')^{i-1} = 1/p'',$$
which is finite. 

\hfill$\Box$

\bigskip

\newpage


{\bf Remarks:}  
\begin{itemize}
\item[(i)] 
The probability of extinction lies strictly between 0 and 1. Conditional on non-extinction, for all time $t>0$, the probability of explosion before time $t$ and after time $t$ are both strictly positive. In particular, $\EE[M_t]$ is infinite for all $t>0$. For the geometric CF model, the  probability of extinction becomes small as we increase $M_0$ and/or $\alpha$. On the other hand, for any value of  $M_0$ and any $\alpha$, we can make $\mu$ large enough so that the probability $M_t$ hits zero is as close to 1 as we wish.

\item[(ii)]
Theorem~\ref{thm2} holds also in the case  where $\mu$ (the extinction rate) changes its value a finite number of times with $t$ (or, more generally, if $\mu$ is a time-variable function, which is uniformly bounded above by some constant $\mu_*$). For example, consider the particular case where $\mu$ undergoes a discrete jump (eg. goes from a small value $\mu$ to a larger value, say $\mu_1$ at time $t_1$). 
In that case, the probability of $M_t$ hitting zero will increase;  we can model this precisely by simply taking the constant rate setting in the CF model with the extinction rate $\mu_1$ and the starting time at $t_1$ (rather than 0) and taking the initial population size at $M_{t_1}$ (rather than $M_0$). The only difference here from the usual CF model is that the initial population size is now a random variable ($M_{t_1}$ rather than $M_0$).

\item[(iii)]  \blue{A simple upper bound on the probability of extinction of $M_t$ in the geometric CF model is  $\min\{1, \left(\frac{\mu}{P\alpha}\right)^{M_0}\}.$  This follows from a standard coupling argument based on two observations: (i)
$(1+\alpha)^n -1 \geq \alpha n$ for all $n\geq 1$ and so the birth rate in a geometric CF model with $n$ items is at least  $\lambda n$ where $\lambda  = P \alpha$, and (ii) it is a classic result (see e.g. \cite{allen})
that a linear birth-death process  with birth and death rates $\lambda$ and $\mu$ (respectively) and starting with $M_0$ individuals has extinction probability $\min\{1, \left(\frac{\mu}{\lambda}\right)^{M_0}\}$.
}
\end{itemize}

\bigskip

Theorem~\ref{thm2} shows that the expected value of $T$ (time to extinction or explosion) is finite and uniformly bounded; however, it does not give an explicit description of it.  Proposition~\ref{prow} below does this;  its proof is given in the Appendix.  \blue{The calculation of the variance of $T$ is more involved, and we have not considered this further in the current paper.}

\begin{proposition}
\label{prow}

Consider the CF model  $M_t$ with $M_0=n$. Let  $e_n =\EE[T^{(n)}]$ be the expected value of  the time until $M_t$ either becomes extinct or explodes. Then:
\begin{itemize}
\item[(i)] For all $n\geq 0$, $e_n = \frac{1}{\lambda'_n} + e_{n-1}\gamma_n  + e_{n+1} (1-\gamma_n)$, with $e_0=0$,
where $\lambda'_n$ and $\gamma_n$ are as in (\ref{lamu}).

\item[(ii)] $e_n = \lim_{N \rightarrow \infty} e^{(N)}_n$, where $e^{(N)}_n$ is the solution to the finite (and invertible)  tridiagonal system of linear equations given by:
$$e_n^{(N)}-e_{n-1}^{(N)}\gamma_n  -e_{n+1}^{(N)} (1-\gamma_n) = \frac{1}{\lambda'_n} $$
for $1\leq n \leq N-1$,
with the boundary conditions: $e_0=e_N=0.$
\end{itemize}
\end{proposition}
We can write this recursion more compactly as a vector equation:
$$A_N {\bf e} = {\bf u},$$
where ${\bf u}=[u_n]$ is the column vector with $u_n =  \frac{1}{\lambda'_n}$, $A_N$ is a $(N-1) \times (N-1)$ tri-diagonal matrix, and ${\bf e}$ is the column vector $[e^{(N)}_n]$ (for $n=1, \ldots, N-1$). Notice that $A_N$ has entries $+1$ down the diagonal, with all its off-diagonal entries being negative and  each row sum equaling 0.

\section{Simulation}  \label{sec:simulation}

We have implemented two discrete-time numerical simulations of the model, one deterministic and the other stochastic. Because in any simulation of the model the terms $\binom{M_t}{i}$ quickly grow out of hand, we have chosen to put an upper limit $K$ on how many goods can potentially be combined into new goods. For the sake of numerical simplicity, we have set $K=4$. However, as the results below show, this is not a severe restriction. As in the numerical example in the previous section, we use $\alpha_i = P \alpha^i$ for some given value of $\alpha$, and $P \in (0, 1]$.

The deterministic model is implemented as follows:

\begin{equation}
\label{wimdet}
M_{t+1} = \blue{(1-\mu)}M_t +  P \cdot \sum_{i=1}^K \alpha^i \binom{M_t}{i} 
\end{equation}

Note that this version of the model is a variation of the one given in \cite{kop18}, since the summation term has a fixed upper bound. Moreover, the values of $M_t$ can be non-integer by applying the usual extension of the definition of $\binom{x}{i}$ to allow $x$ to take non-integer values (at the cost of losing the combinatorial meaning of this term).

A stochastic version of the model\footnote{Similar to, but not exactly identical to the pure-birth CF model due to the use of discrete time and the upper bound value $K$.} is implemented as follows:
\begin{enumerate}
\item Start with an initial number of items $M_0$ at $t=0$.
  \blue{\item Draw a random number $u$ from a Poisson distribution with mean $\mu M_t$. Set $M_t = M_t - u$.}
  \item For $i=1,\ldots,K$, calculate the expected number of new items resulting from a combination of $i$ `parents', as $s_i = P \times \alpha^i \times \binom{M_t}{i}$.
  \item For $i=1,\ldots,K$, draw a random number $r_i$ from a Poisson distribution with mean $s_i$. This gives the actual number of new items $r_i$ resulting from a combination of $i$ parents.
  \item Set $M_{t+1} = M_t + \sum_{i=1}^K r_i$.
  \item Set $t = t+1$.
  \item If $M_t < {\bf M}$, go to Step 2.
\end{enumerate}

\blue{There are two reasons why we have implemented the model as a discrete-time process rather than a continuous-time one. First, computationally it is much faster to execute, while statistically it provides the same results. And second, it is easier to keep track of which goods produce which new goods. This allows us to study other properties of the model, such as descent distributions. We show some preliminary results on this below.}

\begin{figure}[htb]
\centering
\includegraphics[scale=0.6]{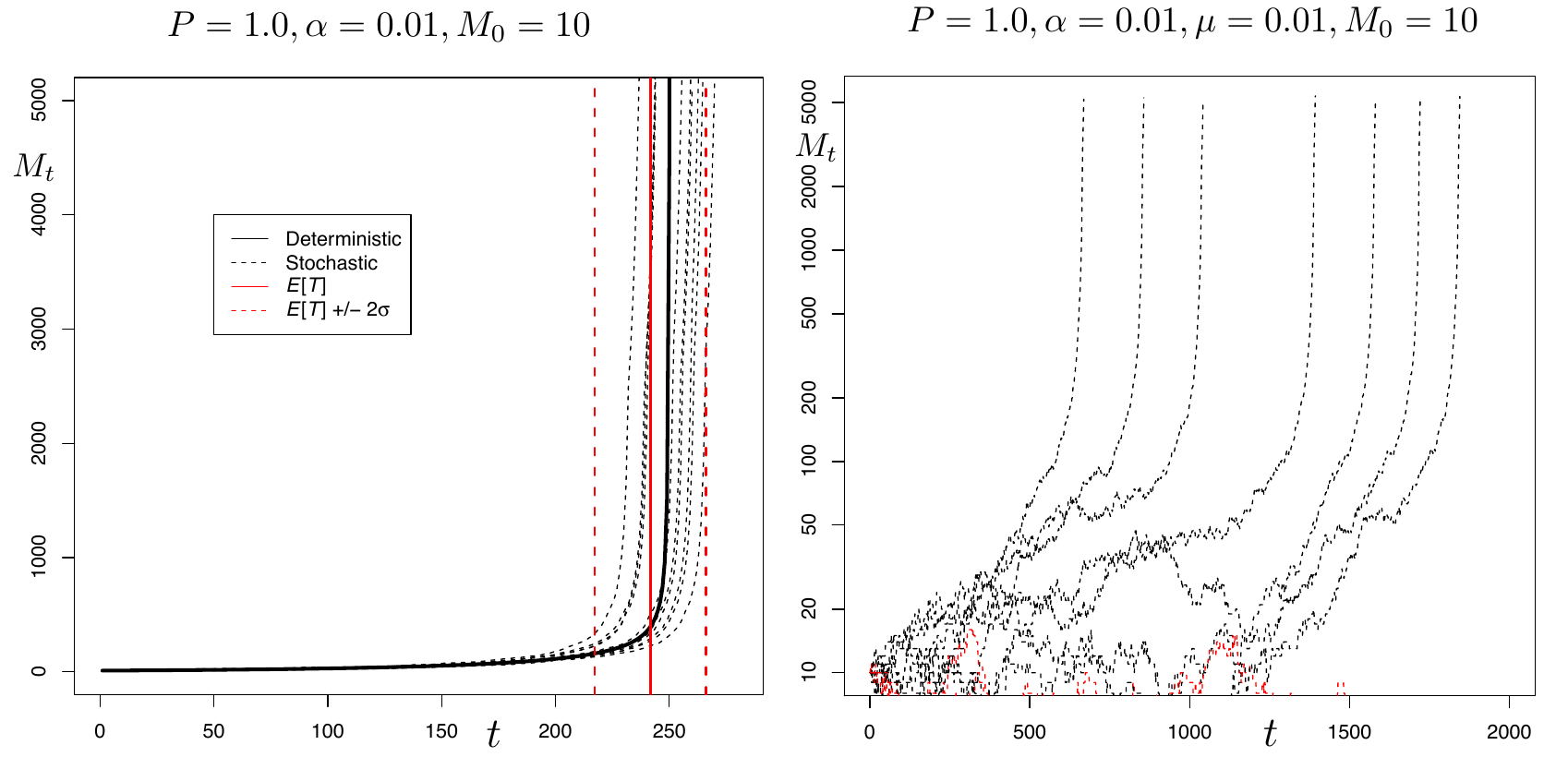}
\caption{{\em Left:} The results of the numerical simulations of the model for both the deterministic version (solid black curve) and the stochastic version (dashed black curves) \blue{with no extinction}. The solid red line represents the theoretically calculated mean time to infinity $\EE[T]$, with the dashed red lines representing $\pm 2 \sigma$. \blue{{\em Right:} Numerical simulations using the stochastic version of the model but allowing extinction. The graph shows the result of 10 runs, 7 of which explode (shown in black) and 3 go extinct (shown in red).}}
\label{fig1}
\end{figure}

First, we evaluate the results of Theorem \ref{mainthm} by running both the deterministic and the stochastic simulation models without extinction, i.e., with $\mu=0$. As in the numerical example above, we use $M_0=10$, $P=1.0$, and $\alpha=0.01$, and run the simulations until a number of goods ${\bf M}=5000$ has been reached. We ran the deterministic model once and the stochastic model 10 times. The results are presented in Fig. \ref{fig1} (left).

The solid black curve results from the deterministic simulation model, while the dashed black curves represent the different runs from the stochastic simulation model. The solid red line is the theoretically calculated value for the mean time to infinity $\EE[T]=241.73$. The dashed red lines represent plus or minus two standard deviations, where the theoretically calculated standard deviation $\sigma=12.26$.

Note that the theoretically calculated mean $\EE[T]$ is slightly smaller than the one resulting from the simulation model. This is partly explained by the fact that the simulation model uses an upper limit $K=4$ on the number of goods that can be combined to produce new goods. This will result in a slightly smaller rate of growth in $M_t$, and thus a slightly larger mean time to infinity. If we take the value ${\bf M}=5000$ to represent `infinity' in the simulation model, then the observed mean is about 250, with a standard deviation of 11.17. These values agree well with the theoretically calculated values, despite the upper limit $K=4$ used in the simulation models. For larger values of $K$, the agreement will be even better.

\blue{Next, we evaluate Theorem \ref{thm2} by running the stochastic simulation model with extinction, setting $\mu=0.01$ (and using the same values for the other parameters as before). Figure \ref{fig1} (right) shows the result of 10 runs, with the vertical axis on a log-scale to clearly show the fluctuations due to extinction. Out of these 10 runs, 7 eventually lead to an explosion (shown in black), while 3 lead to extinction (shown in red). Note that once a number of goods of around $M_t=100$ is reached, explosion follows very quickly, but until then it could go either way.}


\blue{With the stochastic implementation we can also investigate other types of behaviours that follow from the formal model. For example, we can keep track of the ``descendants'' of each good. In particular, each time a new item is produced from a combination of $i$ existing items, then this new item is regarded as a descendant of each of the $i$ items that produced it, as well as a descendant of all the earlier items that have $i$ as an ancestor. This gives rise to a {\it descent distribution} which describes the proportion of items having $0,1,2, \ldots $ descendants.}

\begin{figure}[htbp]
\centering
\includegraphics[scale=0.5]{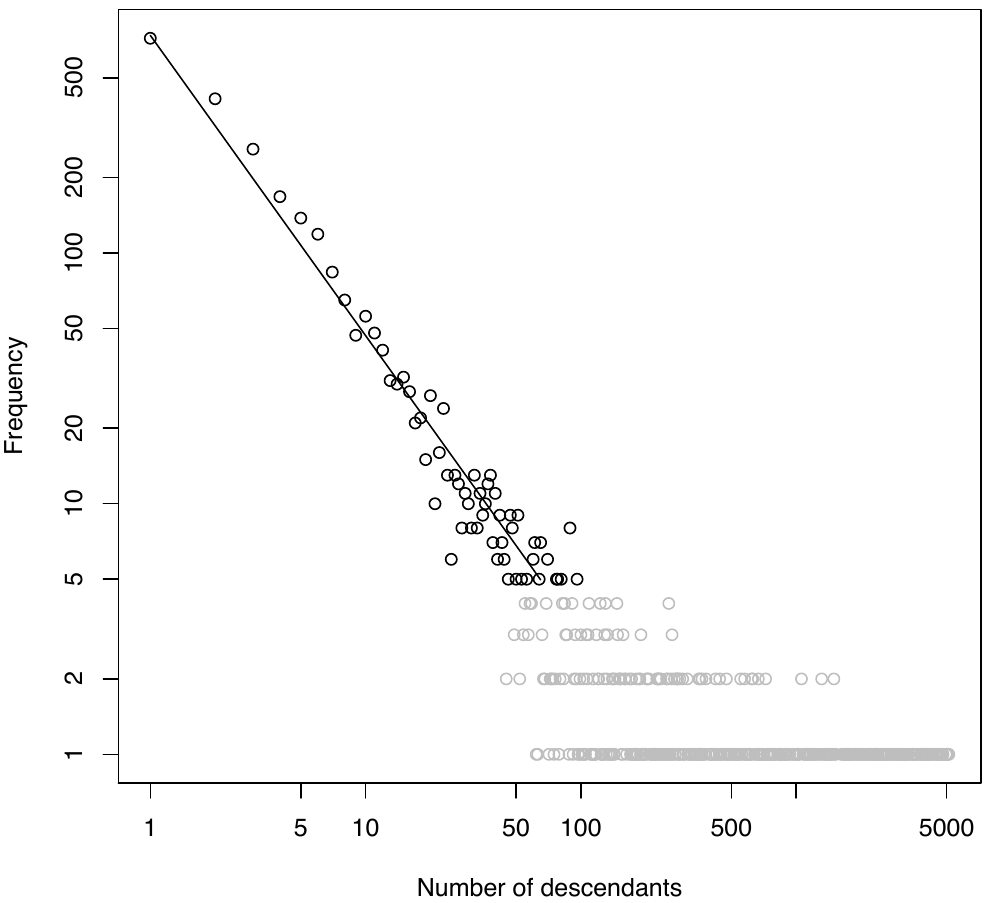}
\caption{\blue{A histogram (in a log-log plot) of the descent distribution of one particular run of the stochastic model. Grey circles represent frequencies of less than 5. The straight line is a regression fit to the black circles.}}
\label{fig:hist}
\end{figure}

\blue{Fig.~\ref{fig:hist} shows such a descent distribution (in a log-log plot) from one particular run of the stochastic model, using the same parameter values as above (but without extinction). This distribution is shown as a histogram, with on the horizontal axis the possible number of descendants and on the vertical axis the number of items that have a given number of descendants.}

\begin{figure}[htbp]
\centering
\includegraphics[scale=0.45]{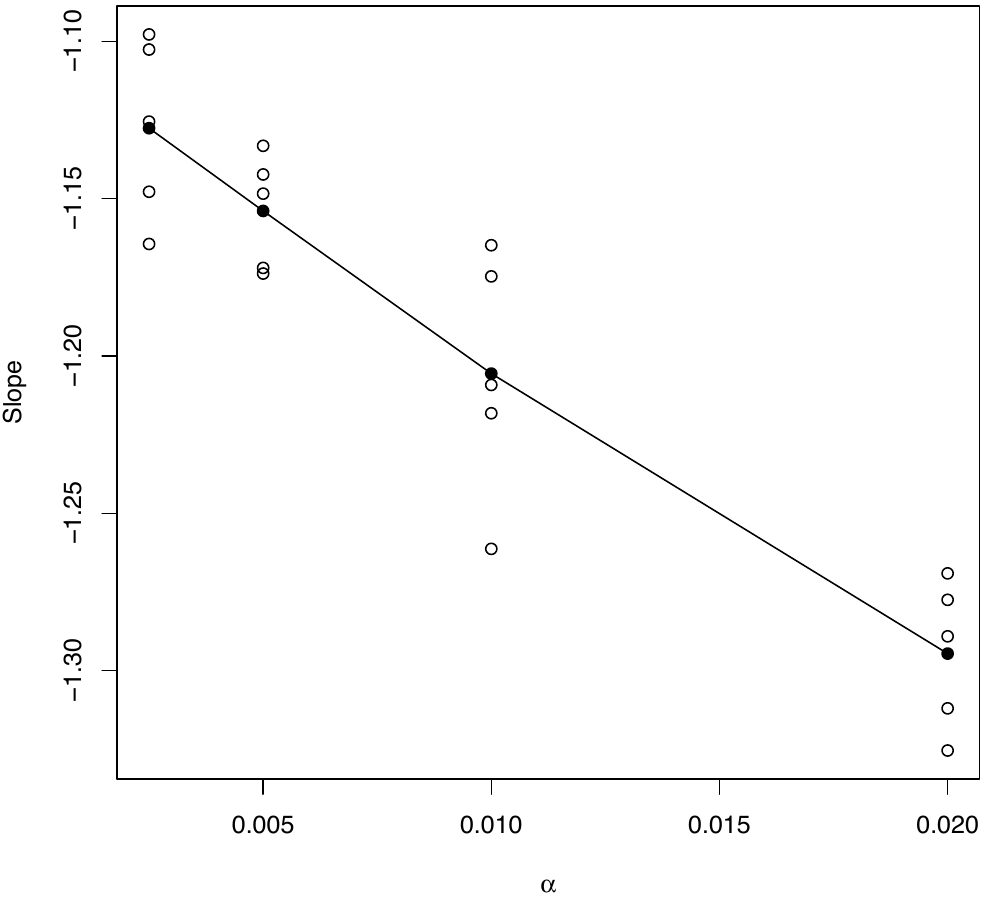}
\includegraphics[scale=0.45]{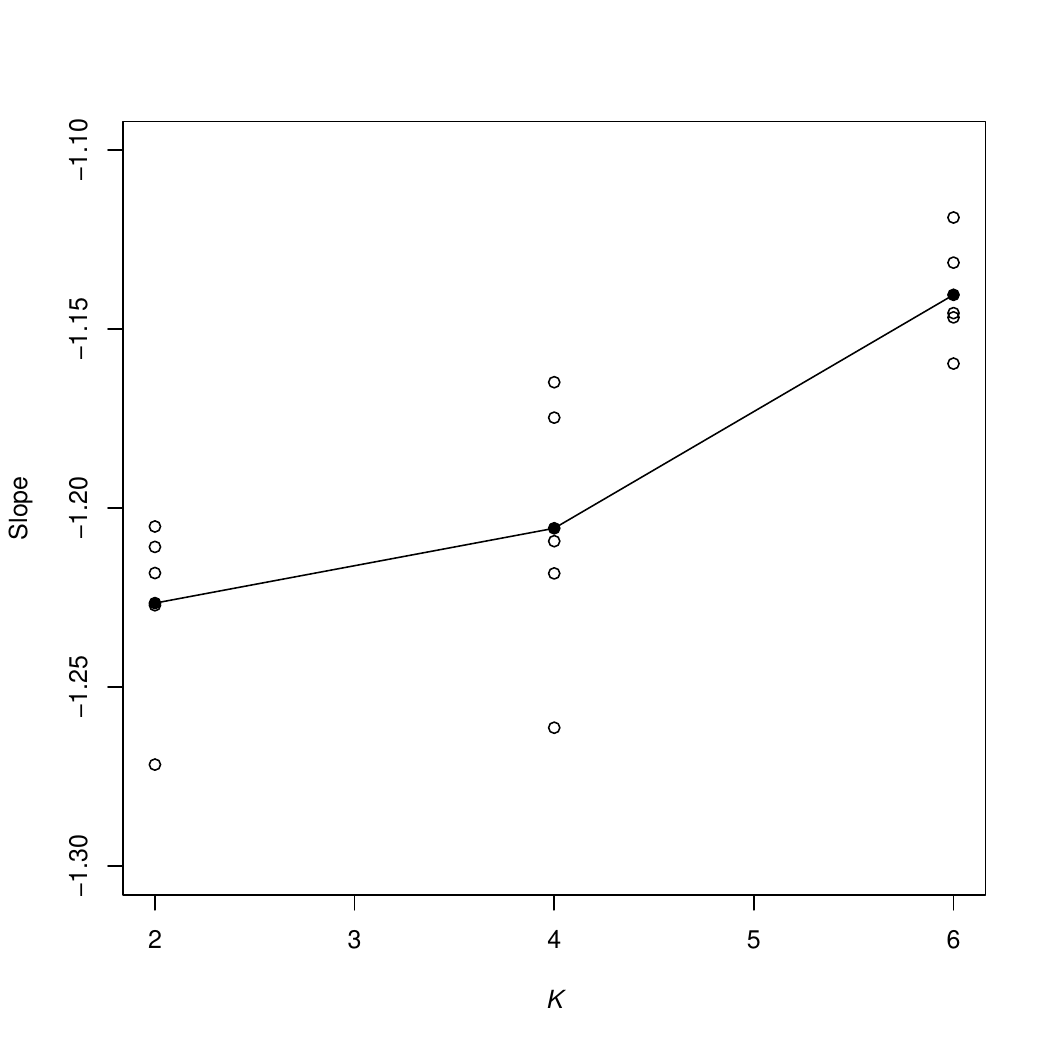}
\caption{\blue{Dependence of the slope of the power law on the model parameters $\alpha$ (left) and $K$ (right).}}
\label{fig:slope}
\end{figure}

\blue{The grey circles in the plot represent frequencies of less than 5. Ignoring those observations, and performing a regression analysis on the data represented by the black circles results in a power law (represented by the straight line) with a slope of -1.198, with a good fit ($R^2=0.93$). The exact slope of the power law depends on the model parameters. Fig. \ref{fig:slope} shows some preliminary results on how the slope depends, in particular, on the parameters $\alpha$ and $K$. Open circles represent individual runs, black circles connected by lines represent averages over these individual runs.}

\blue{Power laws have been observed in, for example, patent data \citep{Youn:15}. Although these authors did not look at descent distributions, they do argue that patent data is a good proxy for technological innovations. It will be interesting to derive descent distributions from such data, which could then be compared to the results from our simulation model. We hope to do this comparison in future work.}


\section{Discussion}

We have formalised and analysed a stochastic model (the CF model) representing a simple cumulative combinatorial growth process of the number of `goods' $M_t$ over time. Our results establish that if the extinction rate $\mu$ is equal to zero in this model, $M_t$ initially grows very slowly, followed by a rapid burst of growth, reaching infinity in finite time $T$ with probability 1.0. We derived a theoretical mean $\EE[T]$ and variance for this time to infinity $T$. Our simulation results fit this growth process and the theoretical calculations very well. If $\mu$ is strictly positive,  the process either becomes extinct or explodes in finite time (the probability of each scenario depends on
the size of the initial population, together with how large $\mu$ is relative to the other parameters). 

We suggest that our model describes the characteristic hockey-stick pattern of initially slow growth then 
rapid explosion in the cumulative technological evolution of humans. The diversity of tools since 2.6 million years ago to the billions of goods at present appears to fit this pattern, at least qualitatively. As a recent proxy for such technological evolution, we can consider global gross domestic product (GDP) over the past 2000 years \citep{kop18}. 
\blue{The notion that the variety, complexity and sophistication of products produced by a country was developed more formally into an `economic complexity index' in \cite{hid}, and has been shown to be a good predictor of GDP per capita growth.}
  
\begin{figure}[htb]
\centering
\includegraphics[scale=0.6]{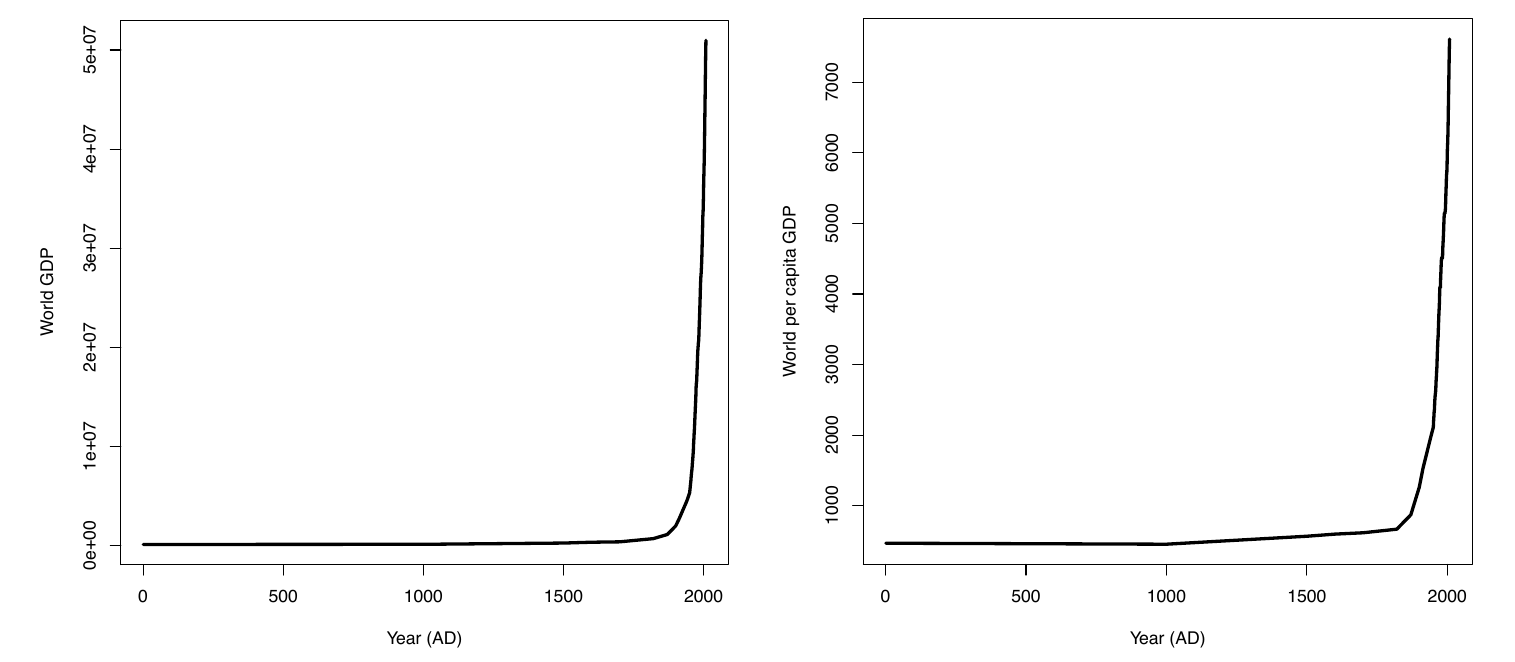}
\caption{World GDP (in 1990 international dollars) over the past two millennia,\blue{ both in absolute value (Left) and per Capita (Right). World DPP Data source: Angus Maddison, {\tt http://www.ggdc.net/maddison/oriindex.htm} (under `Historical Statistics')}.}
\label{fig2}
\end{figure}

Here data are presented \blue{on the left-hand side of} Fig.~\ref{fig2}. Global GDP grew very slowly for most of the past two millennia, until about 1850 (the time of the Industrial Revolution) when it suddenly shot upwards. \blue{One simple explanation for the characteristic shape of this graph is that population growth has also experienced a rapid increase near the present. However, if one takes the ratio of GDP per capita (shown on the right of Fig.~\ref{fig2}) the shape of the resulting curve still maintains a similar overall shape, consistent with the predictions of the type of model described here.}

It should be possible to fit the parameters of the CF model (in particular, $P$ and the $\alpha_i$ parameters) to this actual economic data \citep{kop18}.
One might alternatively explain the shape of the curve in Fig.~\ref{fig2}  as arising from a standard birth--death model, 
where the birth rate was initially low and constant, then rose over a relatively short period of time (100--200 years) as a result of factors that changed the world economy (e.g. mass production, international travel and transport, computing, etc). However, this {\em extrinsic} explanation can also be modelled within the context of our model of {\em intrinsic} growth,  since GDP is related to the  complexity of ideas, products, and processes, which arose through the sequential combination of existing ideas, products and processes over time. 
\blue{We note, however, that GDP can also be quite a course measure of the complexity of some processes; for example, oil extraction in the middle east using relatively low-level technology has a major impact on  GDP in that region.}

There may be other data that could be described directly and quantitatively by our model, such as the diversity of stone tool technology, or the historical records of patents.  Within the former context, the  CF model can  be interpreted as follows.

Let $M_t$ be the most complex good introduced in the economy at time $t$. New, more complex tools can arise from the complex tools already available. The process then describes 
\blue{the increasing complexity and diversity} of these goods into a simultaneous `tool kit' of simple and ever more complex tools. {\it Australopithecus} had perhaps a dozen very simple stone tools. Cro Magnon had hundreds ranging from needles to spear throwers. We have billions ranging from needles to the International Space Station.



All these processes have in common that over time they expand into the `adjacent possible', with new `things' enabling the emergence of even more new `things', in a combinatorial manner. We propose our CF model as a mathematical formalisation of such processes. Here, we have shown the main properties of this model, both theoretically and through simulations. These initial results are encouraging, and we hope to explore the correspondences with some of the above mentioned real-world processes in the future.


\section*{Acknowledgements}
\blue{We  thank an anonymous reviewer for a number of very helpful comments and suggestions concerning an earlier version of this manuscript.}
MS thanks the (former) Allan Wilson Centre for helping fund this research.  WH thanks the Complexity Institute of Nanyang Technological University, Singapore, for support in the form of a fellowship. SK thanks Roger Koppl, Abigail Devereaux, and Jim Herriot.

\section*{Author contributions}
MS: mathematical formulation and analysis; WH: simulation implementation and analysis; SK: formulation of the central equation and broader discussion. All three authors contributed towards the writing of the paper. 



\bibliographystyle{model2-names}
\bibliography{TAP_elsevier_REVISION}

\section{Appendix: Additional proofs}

\noindent {\em Proof of Claim 1 when $\alpha_2>0$ and $\alpha_k=0$ for all $k>2$.}

Our proof will rely on the following lemma. 
It simply asserts that a discrete-time random walk on the integers, starting at zero and with the probability of taking a step to the right being uniformly greater than the probability of
taking a step to the left, has a strictly positive probability of never returning to its starting position. This is a standard result from Markov chain theory.

\begin{lemma}
\label{lem1}
Let $Y_1, Y_2, \ldots, Y_j, \ldots,$ be a sequence of independent random variables taking values in the set $\{+1, -1\}$ with $\PP(Y_i=+1) > c > \frac{1}{2}$ for all $i$. 
Then: $$\PP\left(\forall j \geq 1, \sum_{i=1}^j Y_i >0\right) >0.$$
\end{lemma}

We use this lemma to justify Claim 1 by applying  a coupling argument.
First, select $\beta>\mu$ and then select a sufficiently large value of $m$ such that for all $n \geq m$, we have:
\begin{equation}
\label{eqp}
\alpha_k \binom{n}{2} \geq 2\beta n.
\end{equation}

The process $M_t$ conditioned on $M_0=n$ is stochastically identical to the following
stochastic process $M'_t$. 
Let $ B^{(n)}_t$ be the pure-birth CF process that is initiated at time $t \geq 0$ with $B_t = n$ and which has 
its $\alpha$ values equal to exactly one half of the $\alpha$ values of $M_t$ and with extinction rate $0$. 

Similarly, let $D^{(n)}_t$ be the CF process that starts at time $t \geq 0$ with $B_t = n$ and has  its
$\alpha$ values equal to exactly one half of the $\alpha$ values of $M_t$,  and with extinction rate $\mu$.

Starting at $t=0$, the process $M'_t$ is obtained by running  $B^{(n)}(0)$ and $D^{(n)}(0)$ simultaneously and independently, and at the first time $t'$ that one of these processes changes from $n$ to $n'  \in \{n-1, n+1\}$, one continues the process by running $B^{(n')}(t')$ and $D^{(n')}(t')$ independently.  This  process is then repeated as time proceeds, inorder to give a sequence of values $n, n', \ldots,$ at times
$t=0, t', \ldots.$ In this way, the resulting process $M'_t$ is then stochastically identical to $M_t$. 

Now, the probability that the sum of all the cumulative changes under the $D$ process is strictly positive is some value $p>0$,  since $D$ is dominated by a linear birth--death process with birth rate $\beta>\mu$ and so we may apply Lemma~\ref{lem1} above. 
Thus the process $M'_t$ remains always greater or equal to $n$ and so the probability that
$B^{(n')}(*)$ is called in place of  $D^{(n')}(*)$ always remains at least $\frac{1}{2}$ by Inequality~(\ref{eqp}).
Finally, $D^{(n)}(*)$ represents a pure-birth process that explodes at a finite time, with probability 1.
Claim 1
now follows.

\bigskip

\noindent {\em Proof of Proposition~\ref{prow}}
We begin with a lemma.

\begin{lemma}
\label{base}
Consider the CF model ($M_t$). For $N \in \{1,2, 3, \ldots, \}$, let $E_N(t)$ be the event that $M_{t'} =0$ or $M_{t'} \geq N$ for at least one value of $t'\leq t$.
\begin{itemize}
\item[(i)] Let $E(t)$ be the event that $M_t$ has either reached 0 or exploded by time $t$. Then:
$$\PP(E(t)) = \lim_{N \rightarrow \infty} \PP(E_N(t)).$$
\item[(ii)]
Let $T$ be the time until $M_t$ has either reached 0 or exploded and let $T'_N$ be the first time $t$ at which either $M_{t'}=0$ or $M_{t'} = N$. Then $\EE[T] = \lim_{N \rightarrow \infty} \EE[T'_N]$.
\end{itemize}
\end{lemma}
{\em Proof:}

\noindent {\em Part (i):} For fixed $t$, the sequence of events $E_N(t), N\geq 1$ is a nested decreasing sequence (i.e. $E_{N+1}(t) \subseteq E_{N}$) and $E(t) = \bigcap_{N\geq 1} E_N(t)$). The equation in Part (i) of the lemma is now an elementary identity in probability theory.

\noindent {\em Part (ii):} From Part (i):
$$\PP(T>t) = (1-\PP(E(t)) = \lim_{N \rightarrow \infty}(1- \PP(E_N(t)) = \lim_{N \rightarrow \infty} \PP(T'_N>t).$$
Thus:
$$\EE[T] = \int_{0}^\infty \PP(T>t) dt = \int_{0}^\infty  \lim_{N \rightarrow \infty} \PP(T'_N>t) dt =   \lim_{N \rightarrow \infty}\int_{0}^\infty \PP(T'_N>t) dt= \lim_{N \rightarrow \infty} \EE[T'_N],$$
where monotonicity allows us to exchange the order of the limit and integration. 
\hfill$\Box$

\bigskip

We return now to the proof of Proposition~\ref{prow}.
Recall the definition of $\lambda'_n$ and $\gamma_n$ from (\ref{lamu}). 
By the law of total expectation, for all $n \geq 1$, we have $T^{(n)}= X_n+T'$ where $X_n$ is the time until the population size of $n$ first changes (up or down by 1) and $T'$ is the time from this new population size to explosion or extinction. Thus $\EE[T^{(n)}] = \EE[X_n]+ \EE[T']$. Since $X_n$ has an exponential distribution with rate $\lambda'_n$, we have $\EE[X_n] = \frac{1}{\lambda'_n}$, and thus:
$$e_n = \frac{1}{\lambda'_n} + \EE[T'].$$
By the law of total expectation, we have:
$$\EE[T'] = \EE[T^{(n-1)}]\gamma_n + \EE[T^{(n+1)}](1-\gamma_n) =e_{n-1} \gamma_n + e_{n+1}(1-\gamma_n).$$
Combining these two equations gives us the following equation. For all $n \geq 1$:
\begin{equation}
\label{eneq}
e_n = \frac{1}{\lambda'_n} + e_{n-1}\gamma_n  + e_{n+1} (1-\gamma_n),
\end{equation}
where the boundary condition is $e_0=0$. 

\bigskip

{\em Part (ii):}
With a view to using Lemma~\ref{base},  consider the modified process $M'_t$ as an absorbing finite-state continuous-time Markov process on the state space $0, 1, \ldots, N$, that has  0 and $N$ as its two absorbing states and has the same transition process as $M_t$ on states $1, 2, \ldots, N-1$.   For $n \in 0, 1, \ldots, N$, let  $e^{(N)}_n$ be the expected time until absorption of $M'_t$ when $M'_0=n$  (by classical Markov process theory, $e^{(N)}_n$ is finite). We have $e^{(N)}_0=e^{(N)}_N = 0$. For 
$n=1, \ldots, N-1$, the same argument used to establish Eqn.~(\ref{eneq}) gives the tridiagonal system:
$$- \gamma_n  e^{(N)}_{n-1} + 1\cdot  e^{(N)}_n  - (1-\gamma_n) e^{(N)}_{n+1} = u_n,$$
where $u_n =  \frac{1}{\lambda'_n}$.  Part (ii) of Lemma~\ref{base} now justifies the limit claim that
$e_n = \lim_{N \rightarrow \infty} e^{(N)}_n$.
\hfill$\Box$

\end{document}